# Materials for Future Calorimeters

Minfang Yeh [a,*] and Ren-Yuan Zhu [b,*]

[a] *Brookhaven National Laboratory, Upton, NY 11973, USA*
[b] *California Institute of Technology, Pasadena, CA 91125, USA*

Marcel Demarteau [c], Sarah Eno [d], David Hitlin [b], Frank Porter [b], Richard Rosero [a], Randy Ruchti [e,+], Jui-Jen Wang [f,+]

[c] *Oak Ridge National Laboratory, Oak Ridge, TN37831, USA*
[d] *University of Maryland, College Park, MD 20742, USA*
[e] *University of Notre Dame, Notre Dame, IN 46556, USA*
[f] *University of Alabama, Tuscaloosa, AL 35487-0324, USA*

## ABSTRACT

Future HEP experiments present stringent challenges to calorimeter materials in radiation tolerance, time response and project cost. The 2019 report of the DOE Basic Research Needs Study on High Energy Physics Detector Research and Development points out three priority research directions for future calorimetry. Following these research directions letters of interest were submitted to the Snowmass organized by the Division of Particles and Fields of the American Physics Society. This report summarizes materials to be developed in the form of inorganic, liquid (oil- and water-based), and plastic scintillators and wavelength shifters to advance HEP calorimetry to face the challenges in radiation hardness, fast timing, and cost-effectiveness. Some of these materials may also find applications for future HEP time-of-flight system, and beyond HEP in nuclear physics, hard X-ray imaging and medical instruments.

Submitted to the Proceedings of the US Community Study on the Future of Particle Physics (Snowmass 2021)

**Editors: yeh@bnl.gov and zhu@caltech.edu*
*+Primary contributors*

## CURRENT STATUS AND PROPOSED PROJECTS

The main mission of calorimetry in HEP experiments is to conduct high-precision measurements in energy, position and direction for electrons, photons, and jets. Current and planned projects can be found in the calorimetry sections of the Review of Particle Physics [1]. Homogeneous electromagnetic (EM) calorimetry provides the best energy resolution for electrons and photons with concerns of high cost and radiation hardness in a severe radiation environment. Sampling calorimetry costs less and is easy to achieve a better radiation hardness because of its reduced signal path length, but with concerns of poor energy resolution limited by the sampling fraction.

The DOE Basic Research Needs (BRN) Study on High Energy Physics Detector Research and Development [2] report points out three priority research directions (PRD) for future HEP calorimetry.

1. Enhance calorimetry energy resolution for precision electroweak mass and missing energy measurements.
2. Advance calorimetry with spatial and timing resolution and radiation hardness to master high-rate environments.
3. Develop ultrafast media to improve background rejection in calorimeters and improve particle identification.

Letters of interest (LOIs) were submitted to the 2021 Snowmass study to address R&Ds directly related to these BRN calorimetry PRDs. These LOIs were also presented in the CPAD Instrumentation Frontier Workshop in March 2021 [3].

To meet the severe radiation environment expected at the HL-LHC and the proposed FCC-hh [4], where up to 500 Grad and $5\times10^{18}$ $n_{eq}/cm^2$ of one MeV equivalent neutron fluence are expected by the forward calorimeter, the RADiCAL consortium [5,6] proposed an ultracompact, radiation hard, and fast-timing shashlik calorimetry consisting of W plates interleaved with radiation hard LYSO:Ce (Cerium-doped Lutetium Yttrium Orthosilicate) crystal or LuAG:Ce (Cerium-doped Lutetium Aluminum Garnet) ceramic plates and readout by radiation hard wavelength shifters (WLS) with the goal to develop a precision electromagnetic calorimeter (ECAL) that has an energy resolution of 15%/√E for electrons and photon at future hadron colliding beam experiments. Figure 1 shows two schematics for a RADiCAL module, scaled to the Moliére Radius of EM showers in a sampling ECAL. Novel inorganic scintillators and wavelength shifters with high density, bright and fast light, and excellent radiation hardness are required.

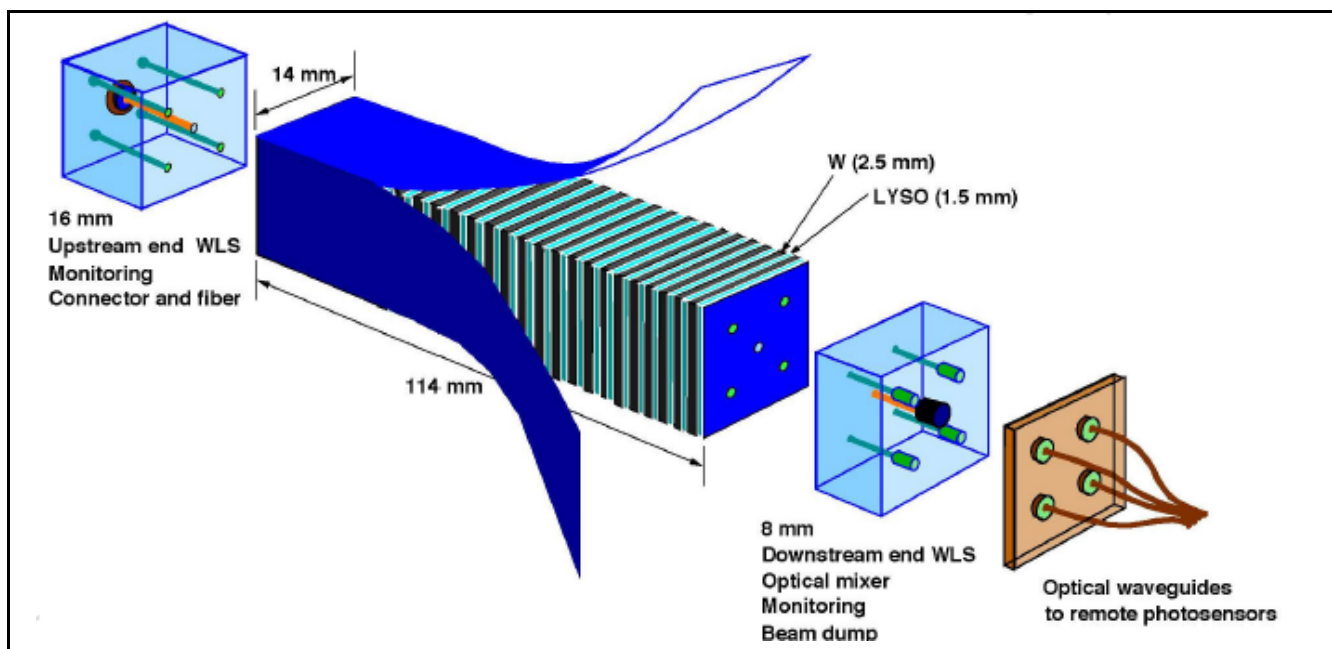

**Figure 1a.** RADiCAL Module Schematic. Module is scaled to the Moliére Radius of EM showers in a W/LYSO:Ce sampling calorimeter. Beam enters from upper left and exits lower right in this view. Light from the scintillation tiles is wave-shifted in capillaries or filaments that penetrate through the full length of the module to photosensors positioned at the ends of the module. If sufficiently rad hard, photosensors can be mounted directly on the upstream and downstream faces of the module. Otherwise, photosensors can be positioned remotely and connected to the module via optical waveguides as indicated in the structure at lower right.

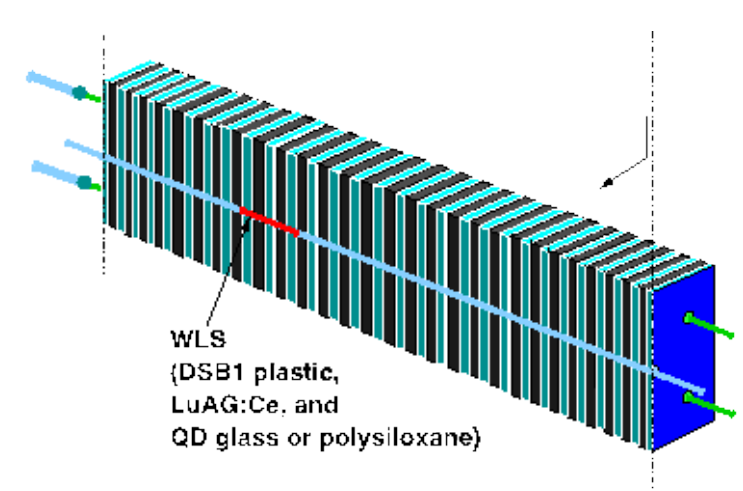

**Figure 1b.** A half-section of a RADiCAL module indicating two energy (E-type) capillaries penetrating the full length of the module and with a timing (T-type) capillary superimposed to show its WLS filament (highlighted in red) positioned in the region of shower max for an EM shower. For the E-type capillaries, the WLS material (not shown) runs the full length of the module. The beam is incident from upper left and exits the module at lower right (indicated by the “blue” surface).

The CalVision consortium [7] proposed a Maximal Information ECAL featured with total absorption, longitudinal segmentation, and dual readout [8] followed by the IDEA dual readout sampling hadron calorimeter (HCAL) [9]. Figure 2 shows the dual readout CalVision crystal ECAL (Left) followed by the IDEA dual readout fiber hadron calorimeter HCAL (Right). Its longitudinal segmentation with heavy inorganic scintillators, such as BGO (Bismuth Germanium Oxide), BSO (Bismuth Silicon Oxide), or PWO (Lead Tungsten Oxide), allows a good capability to implement the particle flow algorithm (PFA), providing good energy, position, and timing resolutions for electrons, photons, and jets for the proposed Higgs factory, e.g., ILC and FCC-ee.

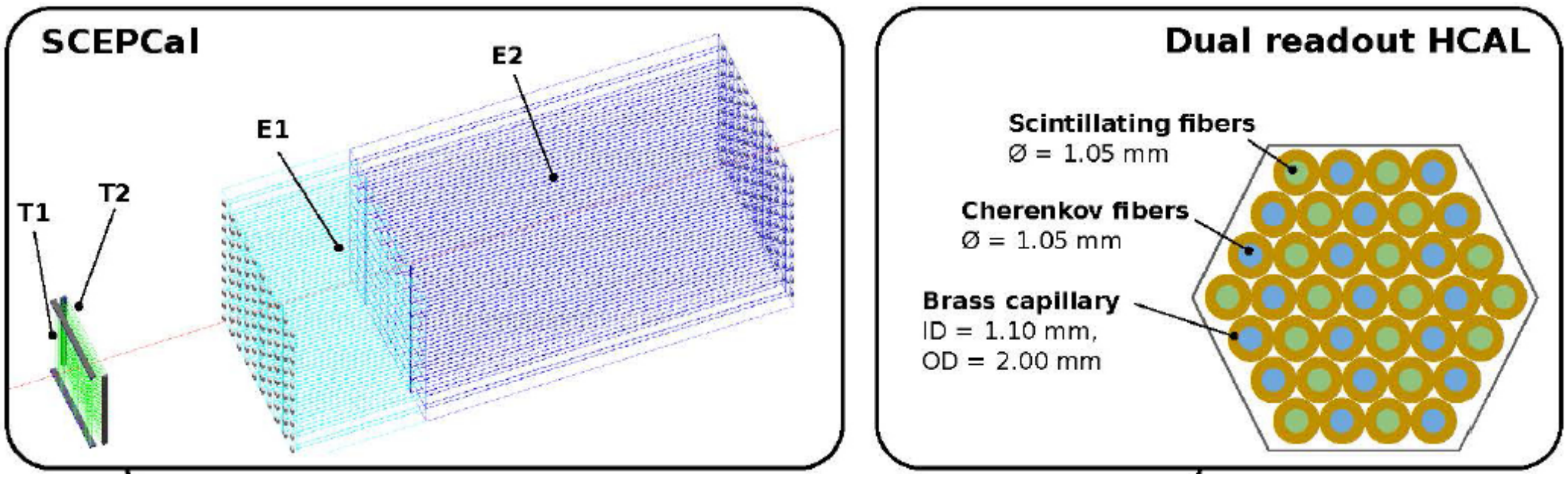

**Figure 2.** The dual readout CalVision crystal ECAL (Left) followed by the IDEA dual readout fiber HCAL (Right).

An even better resolution for jets may be achieved by a total absorption HCAL. Figure 3 (Left) shows a homogeneous hadron calorimeter (HHCAL) concept proposed by the HHCAL consortium [10,6] where a large volume of inorganic scintillators allows event by event corrections with dual readout to recover the missing energies due to nuclei breakup in hadronic jets, providing the best possible jet mass resolution for the proposed Higgs factory. Extensive GEANT simulations conducted at Fermilab [11,12], ANL [13] and CERN [14] show that a better than 20%/√E energy resolution can be achieved for charged pions (Middle and Right) using dual readout of scintillation and Cerenkov light [11,12], or scintillation light in short and long gates [14]. The fine segmentation also allows PFA [13]. Novel cost-effective heavy inorganic crystals and scintillation glasses are crucial for the HHCAL concept.

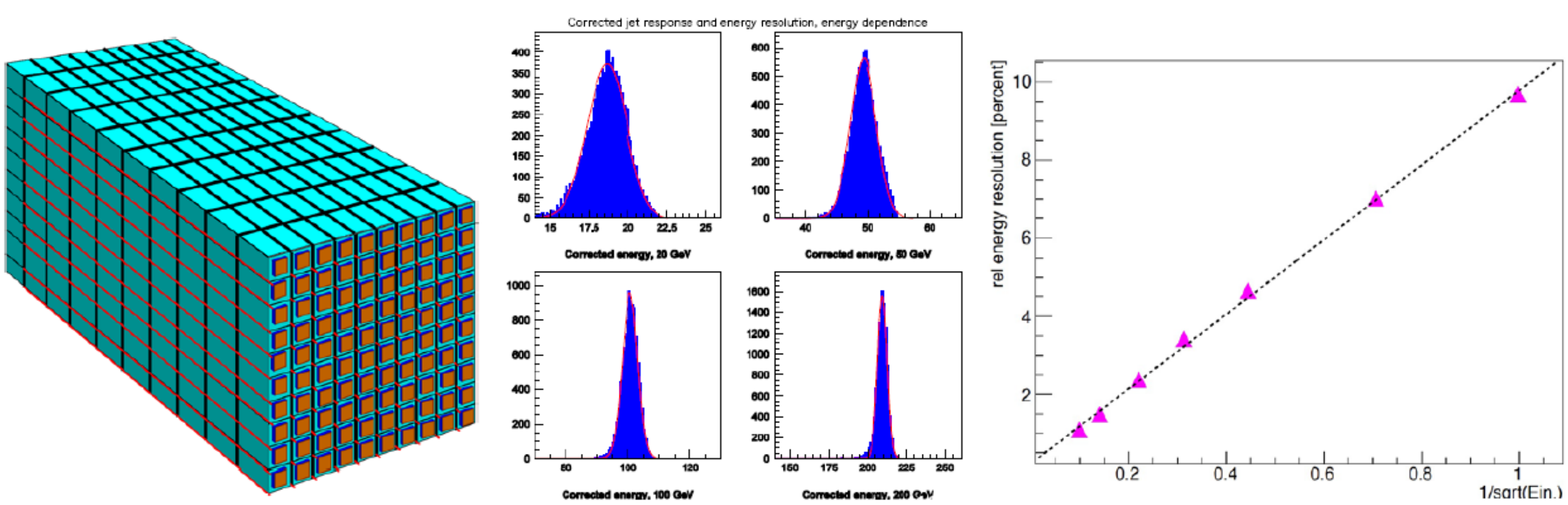

**Figure 3.** The HHCAL concept (Left) promises a 20%/√E energy resolution for charged pions (Middle) and a better than 20% stochastic term (Right) in hadronic energy resolution.

To cope with unprecedented high event rate expected by future HEP experiment at the intensity frontier, the Mu2e-II collaboration has proposed an ultrafast total absorption calorimeter [15,16,17]. Ultrafast heavy inorganic scintillators with sub-nanosecond decay time are preferred for such calorimeter.

An organic scintillator-based calorimeter is low cost in scintillator materials and can be built in any geometric shapes with short decay time, high light yield, and high collection efficiency to satisfy the experimental requirements. However, the ability to reconstruct photons and electrons for electromagnetic calorimeters and to measure charged and neutral hadrons for hadron calorimeters places additional requirements on the energy, position, and shower timing resolutions for organic scintillator calorimeters. An application to utilize liquid scintillator doped with high-Z elements [18] for active materials of calorimeters is proposed. Development of an acrylic-based plastics scintillator is also planned.

## MATERIALS FOR FUTURE CALORIMETRY

Preferred materials for future calorimetry should have the following properties: high density, good optical quality, high light-yield, fast decay time, good radiation hardness and low cost. High density increases stopping power and reduces calorimeter volume, thus the cost. Good optical transmission enhances signal efficacy and thus uniformity. High light-yield improves signal to noise ratio and thus energy, spatial and timing resolution, and reconstruction efficiency. Fast decay time improves timing resolution and the ability to mitigate high event rate. Good radiation hardness is essential for calorimetry survivability, so is crucial to improve the calorimetry stability. Low cost enhances the project affordability and enables the HHCAL concept.

**Table 1.** Optical and scintillation properties of fast and ultrafast inorganic scintillators investigated for HEP TOF system

| | $BaF_2$ | $BaF_2$:Y | ZnO:Ga | YAP:Yb | YAG:Yb | β-$Ga_2O_3$ | LYSO:Ce | LuAG:Ce | YAP:Ce | GAGG:Ce | LuYAP:Ce | YSO:Ce |
|---|---|---|---|---|---|---|---|---|---|---|---|---|
| Density (g/cm$^3$) | 4.89 | 4.89 | 5.67 | 5.35 | 4.56 | 5.94 | 7.4 | 6.76 | 5.35 | 6.5 | 7.2[f] | 4.44 |
| Melting points (°C) | 1280 | 1280 | 1975 | 1870 | 1940 | 1725 | 2050 | 2060 | 1870 | 1850 | 1930 | 2070 |
| $X_0$ (cm) | 2.03 | 2.03 | 2.51 | 2.77 | 3.53 | 2.51 | 1.14 | 1.45 | 2.77 | 1.63 | 1.37 | 3.10 |
| $R_M$ (cm) | 3.1 | 3.1 | 2.28 | 2.4 | 2.76 | 2.20 | 2.07 | 2.15 | 2.4 | 2.20 | 2.01 | 2.93 |
| $\lambda_I$ (cm) | 30.7 | 30.7 | 22.2 | 22.4 | 25.2 | 20.9 | 20.9 | 20.6 | 22.4 | 21.5 | 19.5 | 27.8 |
| $Z_{eff}$ | 51.6 | 51.6 | 27.7 | 31.9 | 30 | 28.1 | 64.8 | 60.3 | 31.9 | 51.8 | 58.6 | 33.3 |
| dE/dX (MeV/cm) | 6.52 | 6.52 | 8.42 | 8.05 | 7.01 | 8.82 | 9.55 | 9.22 | 8.05 | 8.96 | 9.82 | 6.57 |
| $\lambda_{peak}$ [a] (nm) | 300<br>220 | 300<br>220 | 380 | 350 | 350 | 380 | 420 | 520 | 370 | 540 | 385 | 420 |
| Refractive Index[b] | 1.50 | 1.50 | 2.1 | 1.96 | 1.87 | 1.97 | 1.82 | 1.84 | 1.96 | 1.92 | 1.94 | 1.78 |
| Normalized Light Yield[a,c] | 42<br>4.8 | 1.7<br>4.8 | 6.6[d] | 0.19[d] | 0.36[d] | 6.5<br>0.5 | 100 | 35[e]<br>48[e] | 9<br>32 | 115 | 16<br>15 | 80 |
| Total Light yield (ph/MeV) | 13,000 | 2,000 | 2,000[d] | 57[d] | 110[d] | 2,100 | 30,000 | 25,000[e] | 12,000 | 34,400 | 10,000 | 24,000 |
| Decay time[a] (ns) | 600<br>0.5 | 600<br>0.5 | <1 | 1.5 | 4 | 148<br>6 | 40 | 820<br>50 | 191<br>25 | 53 | 1485<br>36 | 75 |
| LY in 1$^{st}$ ns (photons/MeV) | 1200 | 1200 | 610[d] | 28[d] | 24[d] | 43 | 740 | 240 | 391 | 640 | 125 | 318 |
| LY in 1$^{st}$ ns/Total LY | 9.2% | 60% | 31% | 49% | 22% | 2.0% | 2.5% | 1.0% | 3.3% | 1.9% | 1.3% | 1.3% |
| 40 keV Att. Leng. (1/e, mm) | 0.106 | 0.106 | 0.407 | 0.314 | 0.439 | 0.394 | 0.185 | 0.251 | 0.314 | 0.319 | 0.214 | 0.334 |

[a] top/bottom row: slow/fast component; [b] at the emission peak; [c] normalized to LYSO:Ce; [d] excited by alpha particles; [e] ceramic with 0.3 Mg at% co-doping; [f] density for composition $Lu_{0.7}Y_{0.3}AlO_3$:Ce

Table 1 lists optical and scintillation properties for selected fast and ultrafast inorganic scintillators investigated recently [6]. $Lu_{2(1-x)}Y_{2x}SiO_5$ or LYSO:Ce and $Lu_3Al_5O_{12}$ or LuAG:Ce show high stopping power, high light output, fast decay time and excellent radiation hardness against ionization doses and hadrons [6]. LYSO:Ce crystals are used to construct a barrel timing layer (BTL) for the CMS upgrade for the HL-LHC, where the attenuation length of scintillation light is required to be longer than 3 m after radiation doses of 5 Mrad from ionizations of $2.5\times10^{13}$ charged hadrons/cm$^2$ and $3\times10^{14}$ 1-MeV equivalent neutrons/cm$^2$ [19]. While LYSO:Ce crystals satisfy such requirement, LuAG:Ce ceramics shows a factor of two better radiation hardness than LYSO:Ce crystals [6]. They both were proposed by the RADiCAL consortium [5] for the HL-LHC and the proposed FCC-hh. Ultrafast inorganic scintillators may also help to break the picosecond timing barrier for time of flight (TOF) system at future collider experiments. Figures of merit for the TOF application are the light yield (LY) as photons per MeV energy deposition and the ratio between the LY and the total light in the 1st nanosecond as shown in Table 1.

WLS capillaries and fiberoptic filaments form the optical “bridges” that connect the scintillation light emission from the scintillator plates to photosensors. The light collection efficiency of a colorimeter thus depends on light propagation (absorption and reemission) between the scintillator and the WLS and their energy response to the quantum efficiency of the photosensors. Ideally if sufficiently rad hard, photosensors could be mounted directly to scintillation plates positioned at shower max to avoid the use of WLS. This configuration can improve precision timing measurement and light collection efficacy. The performance of mixed modular configurations using different inorganic scintillators and photosensor combinations with or without WLS is summarized in Table 2, under investigation by RADiCAL.

**Table 2.** RADiCAL Radiation Hard Scintillator, Wavelength Shifter and Photosensor Combinations

| Scintillator material | Emission Wavelength | Wavelength Shifter | Emission Wavelength | Photosensor Options |
|---|---|---|---|---|
| LYSO:Ce | 425nm | DSB1 | 495nm | SiPM, GaInP, nVLP |
| LYSO:Ce | 425nm | LuAG:Ce | 520nm | SiPM, GaInP, nVLP |
| LYSO:Ce | 425nm | | | SiPM, nVLP |
| LuAG:Ce | 520nm | Quantum Dots | 560-580nm | SiPM, GaInP, nVLP |
| LuAG:Pr | 310nm | pTP, TPB, Flavenol | 360,460,560nm | SiPM, GaInP, nVLP |
| LuAG:Pr | 310nm | | | SiC |
| $CeF_3$ | 330mm | pTP, TPB, Flavenols | 360nm, 460nm, 560nm | SiPM, GaInP, nVLP |
| $CeF_3$ | 330nm | | | SiC |
| $BaF_2$:Y | 220nm | | | Diamond |

**Table 3.** Optical and scintillation properties of candidate inorganic scintillators for CalVision and the HHCAL concept

| | BGO | BSO | PWO | $PbF_2$ | PbFCl | Sapphire:Ti | AFO Glass | DSB:Ce Glass[1] | DSB:Ce,Gd Glass[2,3] | HFG Glass[4] |
|---|---|---|---|---|---|---|---|---|---|---|
| Density (g/cm³) | 7.13 | 6.8 | 8.3 | 7.77 | 7.11 | 3.98 | 4.6 | 3.8 | 4.7 - 5.4 | 5.95 |
| Melting point (°C) | 1050 | 1030 | 1123 | 824 | 608 | 2040 | 980[5] | 1420[6] | 1420[6] | 570 |
| $X_0$ (cm) | 1.12 | 1.15 | 0.89 | 0.94 | 1.05 | 7.02 | 2.96 | 3.36 | 2.14 | 1.74 |
| $R_M$ (cm) | 2.23 | 2.33 | 2.00 | 2.18 | 2.33 | 2.88 | 2.89 | 3.52 | 2.56 | 2.45 |
| $\lambda_I$ (cm) | 22.7 | 23.4 | 20.7 | 22.4 | 24.3 | 24.2 | 26.4 | 32.8 | 24.2 | 23.2 |
| $Z_{eff}$ value | 72.9 | 75.3 | 74.5 | 77.4 | 75.8 | 11.2 | 42.8 | 44.4 | 48.7 | 56.9 |
| dE/dX (MeV/cm) | 8.99 | 8.59 | 10.1 | 9.42 | 8.68 | 6.75 | 6.84 | 5.56 | 7.68 | 8.24 |
| Emission Peak[a] (nm) | 480 | 470 | 425<br>420 | \ | 420 | 300<br>750 | 365 | 440<br>460 | 440<br>460 | 325 |
| Refractive Index[b] | 2.15 | 2.68 | 2.20 | 1.82 | 2.15 | 1.76 | \ | \ | \ | 1.50 |
| LY (ph/MeV)[c] | 7,500 | 1,500 | 130 | \ | 150 | 7,900 | 450 | 3,150 | 2,500 | 150 |
| Decay Time[a] (ns) | 300 | 100 | 30<br>10 | \ | 3 | 300<br>3200 | 40 | 180<br>30 | 120, 400<br>50 | 25<br>8 |
| d(LY)/dT (%/°C)[c] | -0.9 | ? | -2.5 | \ | ? | ? | ? | -0.04 | -0.04 | -0.37 |
| Cost ($/cc) | 6.0 | 7.0 | 7.5 | 6.0 | ? | 0.6? | ? | 2.0 | 2.0? | ? |

a. Top line: slow component, bottom line: fast component.
b. At the wavelength of the emission maximum.
c. At room temperature (20°C).

1. E. Auffray, et al., J. Phys. Conf. Ser. 587, 2015
2. R. W. Novotny, et al., J. Phys. Conf. Ser. 928, 2017
3. V. Dormenev , et al., the ATTRACT Final Conference
4. E. Auffray, et al., NIMA 380 (1996), 524-536
5. R. A. McCauley et al., Trans. Br. Ceram. Soc., 67. 1968
6. I. G. Oehlschlegel, Glastech. Ber. 44, 1971

Low density crystals/glasses

A Higgs factory, such as ILC and FCC-ee, is proposed as the next HEP collider of priority. Calorimeters with good EM and jet resolutions are required for studying all decay channels of the Higgs boson. In addition to PFA-based W-Si high granularity calorimetry, the CalVision consortium proposed a dual readout CalVision crystal ECAL [7] followed by the IDEA fiber HCAL [9]. It promises a 3%/√E resolution for electrons and photons, and a 27%/√E resolution for neutral pions [8] which is much better than 60% achieved so far by sampling hadron calorimetry. An even better resolution for hadronic jets may be achieved by the proposed total absorption HHCAL concept [10], where a better than 20%/√E energy resolution can be achieved for charged pions by using dual readout of scintillation and Cerenkov light [12], or scintillation light in short and long gates [14]. The total absorption in inorganic scintillators allows event by event corrections with dual readout to recover the missing energies due to nuclei breakup in hadronic jets [11]. Because of the huge volume of inorganic scintillators required for such calorimetry, development of cost-effective inorganic scintillators is crucial. The material of choice must be dense, cost-effective, and UV transparent allowing for a clear discrimination between the Cerenkov and scintillation light, such as BGO, BSO and PWO. Table 3 lists optical and scintillation properties of candidate inorganic scintillators for CalVision and the HHCAL concept, where the cost of material per nuclear interaction length ($\lambda_I$) is the figure of merit for such an application.

Inorganic scintillators with core valence transition features with its energy gap between the valence band and the uppermost core band less than the fundamental bandgap, allowing an ultrafast decay time. As an example, $BaF_2$ crystals have an ultrafast cross-luminescence scintillation with 0.5 ns decay time peaked at 220 nm, but also a 600 ns slow component peaked at 300 nm with a much higher intensity. The latter causes pileup in a high-rate environment. The left plot of Figure 4 shows the scintillation pulse shape measured by a photomultiplier (Top, PMT) and a micro channel photomultiplier (Bottom, MCP-PMT) for a $BaF_2$ sample. A decay time of 0.5 ns is observed by the MCP-PMT, but not the PMT, where 1.4 ns decay time is due to the slow response time of the PMT. It is also known that the slow component in $BaF_2$ crystals may be suppressed either by rare earth doping [6] in crystals or by using a solar blind photodetector [16,17]. The middle and right plots of Figure 4 show respectively the X-ray excited emission spectra and light output as a function of integration time for $BaF_2$ and $BaF_2$:Y cylinders of $\Phi 18\times 21$ mm$^3$ with different $Y^{3+}$ doping levels [6]. They show a reduced slow light intensity for increased yttrium doping level, while the intensity of the fast emission is maintained. As listed in Table 1, yttrium doped barium fluoride crystals ($BaF_2$:Y) have the highest light yield in the 1$^{st}$ nanosecond and the highest ratio between the light yield in the 1$^{st}$ nanosecond and the total light yield, which are the figures of merit for a TOF system. An ultrafast $BaF_2$:Y total absorption calorimeter is also considered by the Mu2e-II collaboration [15], and also by the RADiCAL consortium for ultrafast timing.

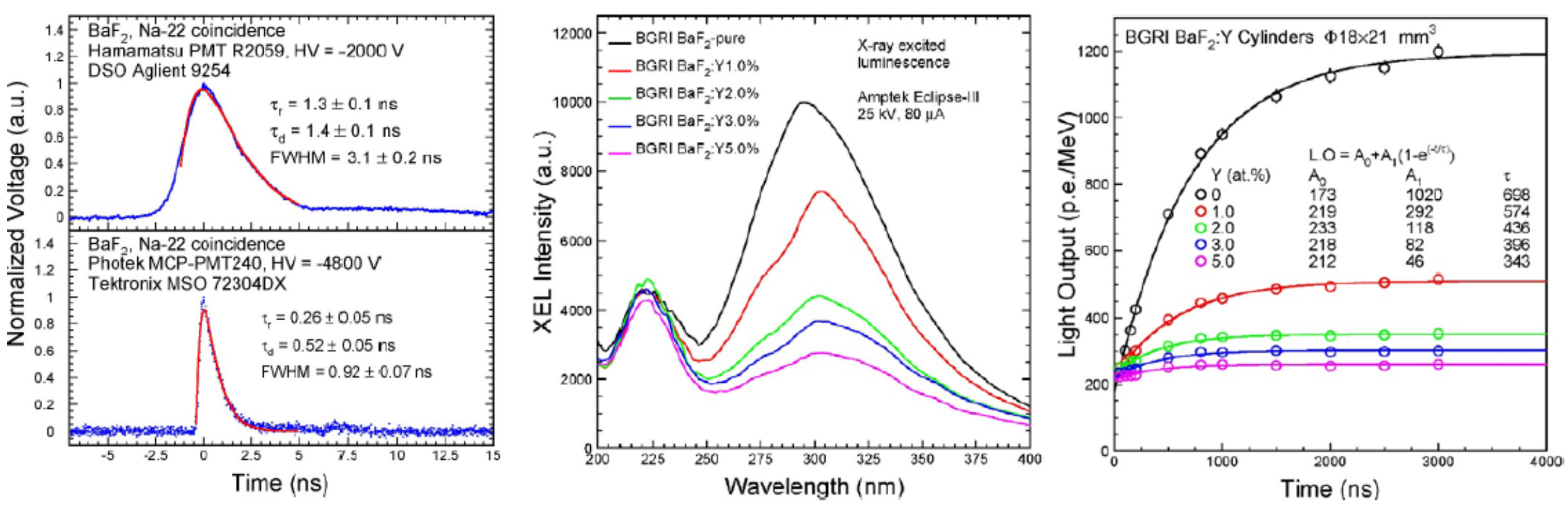

**Figure 4.** Left: The pulse shape measured by a PMT (top) and a MCP (bottom) shows the ultrafast scintillation light component with 0.5 ns decay time for a $BaF_2$ sample. The emission (Middle) and integrated light output (Right) are shown for $BaF_2$ samples with yttrium doping at different levels.

Plastic scintillators are widely used for ionizing radiation detection due to modest cost ($10s per kg) and scale-up availability. The most common type of plastic scintillator is composed of selected fluors and wavelength shifters in a plastic base that is an aromatic compound with delocalized $\pi$-electrons. A detector configuration of plastic scintillators tiled with WLS fibers is a popular choice for sampling calorimeters [20,21,22]. Most plastic scintillators employ polyvinyltoluene (PVT) and polystyrene (PS) as the base resins for fabrication. A new thermoplastics acrylic scintillator aiming to load scintillator materials and high-Z elements directly into acrylic monomers is under development. The mechanically advantageous property, good thermal resistance, and reduced aging effect make the acrylic a promising candidate for constructing large plastic detectors. By using modern SiPMs, a multilayer acrylic detector could further provide excellent position and energy reconstruction. The success of this imrpoved acrylic-based scintillator material has applications in calorimetry and other particle physics fields, such as neutrino and dark matters.

In past decade, liquid scintillator (LS) research has been on the rise on many fronts, such as availability of new scintillator materials, enhanced performance in optical transparency and light-yield, long chemical stability and particularly improved compatibility with tile materials including plastics polymers, such as O-rings and gaskets, which hold the liquids [18]. Another improvement in LS development is the capability of loading metallic elements at high mass percent (i.e., >10w%) from the deployment of several scintillator-based neutrino experiments over the past decade [23]. The newly developed Water-based Liquid scintillator (WbLS) is a low energy-threshold scintillation water that is even capable of loading any target elements at >20w% and still has appreciable scintillation emissions [24]. Figure 5 summarizes the loading elements in liquid scintillators and their usages in particle physics experiments.

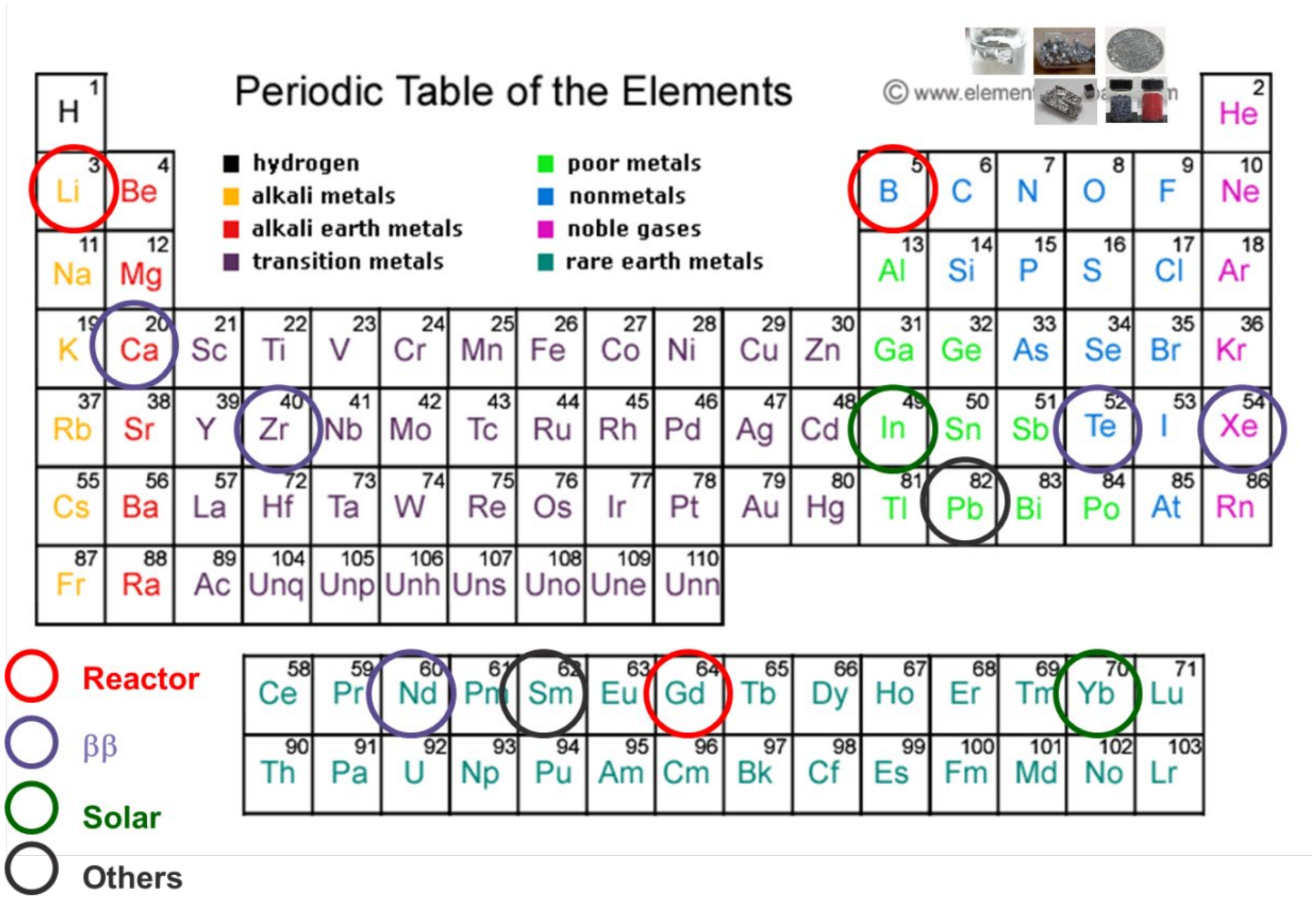

**Figure 5.** Summary of chemical elements loaded in liquid scintillator for physics experiments.

The common features of the new class of liquid scintillators are cost effective (~$3k per ton), environmentally friendly, and less chemical attack on most tile materials. Linear alkyl benzene (LAB), the choice of scintillation solvent for several current neutrino experiments, is an intermediate chemical used for manufacture of household and industrial cleaning detergent. LAB has a high flash point and is considered environmentally friendly in industry. WbLS utilizes industrially available emulsifiers to mix scintillator solvents with water, a hybrid scintillation and Cerenkov detection medium for varied particle detections. It contains more than 50% water and is not combustible under fire hazard. Furthermore, doped scintillator utilizes a target material of choice in liquid scintillator in use for a variety of nuclear and particle physics experiments over the past decades. For antineutrino detection as it utilizes the time-delayed coincidence signature of the positron annihilation and neutron capture resulting from the Inverse Beta Decay (IBD) interaction. The scintillator dopant (gadolinium or lithium) increases the neutron capture cross-section, shortens the capture time, and provides a more distinct signal than the single 2.2 MeV gamma ray emitted after neutron capture on hydrogen. The loading technique can be extended to include most high-Z metallic elements, which could further increase the density of scintillator medium.

Liquid scintillator can be homogeneously deployed in any shape of tiles or containers without geometric limitations and is resistant to radiation dose. Previous studies indicated that conventional liquid scintillators have little decrease of light output with dose [25,26]. The radiation hardness in aging performance of light-yield and optical transmission for LAB and WbLS was also investigated using 201 MeV proton beams that deposited doses of approximately 52 Gy, 300 Gy, and 800 Gy in the tested liquids [27]. The results showed that less than 2% of degradation was observed after ~800 Gy of proton dose on these scintillator materials. Figure 6 presents the Compton light-yield spectra before and after radiation.

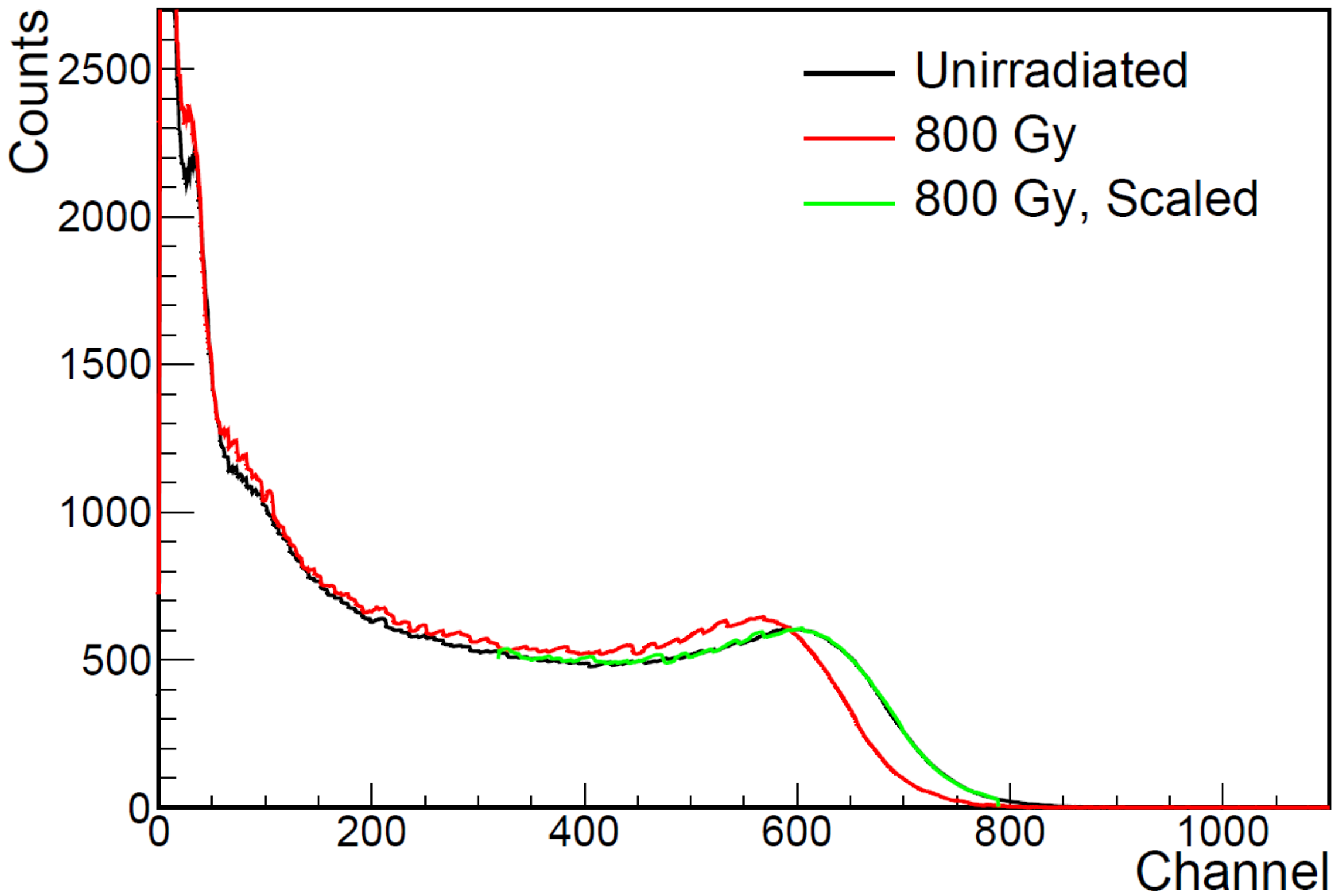

**Figure 6.** An example of the measured Compton electron spectrum for the control (black) and sample of liquid scintillator irradiated by 800 Gy of 201 MeV protons (red). The green trace shows the irradiated spectrum scaled by the amount of light yield loss as determined by the algorithm described in the text. The non-statistical fluctuations are an artifact of the liquid scintillation spectrometer readout.

## SUMMARY

Future HEP experiments at the energy frontier require fast and radiation hard calorimetry. The RADiCAL concept utilizes bright, fast and radiation hard LYSO:Ce crystals and LuAG:Ce ceramics for an ultra-compact, ultra-radiation hard and longitudinally segmented shashlik calorimeter for the HL-LHC and the proposed FCC-hh.

The proposed lepton Higgs factory requires good EM and jet resolutions. The dual readout CalVision crystal ECAL followed by the IDEA fiber HCAL provides an excellent option. Because of its total absorption nature, the HHCAL concept promises the best jet mass resolution. Crucial R&D is to develop cost-effective heavy inorganic scintillators of large volume.

Future HEP experiments at the energy and intensity frontiers require ultrafast calorimetry to mitigate high event rate and break the picosecond timing barrier, where $BaF_2$:Y crystals and solar-blind VUV photodetectors are under development for the proposed Mu2e-II experiment.

Organic-based scintillator calorimeters, including plastics, pure liquid scintillator, and water-based liquid scintillators, have fast pulse response (1-2 ns) with adequate light-yield ($10^4$ photons/MeV). Liquid scintillator and water-based liquid scintillator are less sensitive to radiation damage and have long optical transparency (attn. length of >10m at 450nm) that are capable of deployment in conjunction with any detector compartments in high dose environments. The fabrication and stability of metal-doped liquid scintillators (i.e., gadolinium, lithium, boron) at ton-scale have been demonstrated by the Daya Bay, LZ and PROSPECT experiments. The principal of loading high-Z elements at high mass precent (i.e., lead, indium, tellurium at >10%) was proved by solar (LENS) and double-beta decay (SNO+) scintillator research. The application of WbLS in nonproliferation and multi-physics detection is in progress for WACTHMAN and THEIA. A 30-Ton Demonstrator to prove the WbLS deployment at kiloton scale is under construction. Significant progress towards developing extractants and techniques that will allow the loading of high-Z elements into high flashpoint and less chemical-aggressive scintillator materials has been achieved by different frontiers. Promising scintillator materials identified for calorimetry are linear alkylbenzene (LAB), cyclohexylbenzene (PCH), 1-phenyl-1-xylyl-ethane (PXE) and di-isopropylnaphthalene (DIN). The approach here is to extend the usage of metal-doped (water-based) liquid scintillators, built on the advanced techniques developed from either deployed or developing neutrino and dark matter search, as active materials for sampling calorimeters through measurements of light collection efficiency, uniformity, and radiation hardness at test beams.

## ACKNOLEDGEMENTS

This work is supported by the U.S. Department of Energy under contract number DE-SC0011925, DE-SC0017810, DEAC02-98CH10886.